\documentclass{PoS}

\usepackage{url}

\pdfoutput=1

\title{Towards an interoperable International Lattice Datagrid}

\ShortTitle{Towards an interoperable International Lattice Datagrid}

\author{%
	Paul Coddington, Shunde Zhang\\
	School of Computer Science, University of Adelaide,
        Adelaide, 5005, Australia\\
	E-mail:	\email{paulc@cs.adelaide.edu.au}
		\email{shunde.zhang@adelaide.edu.au}
}

\author{%
	Noriyoshi Ishii, Mitsuhisa Sato\\
	Center for Computational Sciences, University of Tsukuba, Tsukuba, Ibaraki 305-8577, Japan\\
	E-mail:	\email{ishiin@ccs.tsukuba.ac.jp},
		\email{msato@cs.tsukuba.ac.jp}
}

\author{%
	David Melkumyan, \speaker{Dirk Pleiter}\\
	Deutsches Elektronen-Synchrotron DESY, 15738 Zeuthen, Germany\\
	E-mail:	\email{dirk.pleiter@desy.de},
		\email{david.melkumyan@desy.de}
}

\author{%
	George Beckett, Radoslaw Ostrowski\\
	EPCC, School of Physics, University of Edinburgh, Edinburgh EH9 3JZ,
	Scotland\\
	E-mail:	\email{george.beckett@ed.ac.uk}
		\email{radek@epcc.ed.ac.uk}
}

\author{%
	James Simone\\
        Fermi National Accelerator Laboratory, Batavia, Illinois 60510, USA\\
        E-mail:	\email{simone@fnal.gov}
}

\author{%
	Balint Jo\'o, Chip Watson\\
        Jefferson Lab, Newport News, VA 23606, USA\\
        E-mail:	\email{bjoo@jlab.org},
        	\email{watson@jlab.org}
}

\author{ILDG Middleware Working Group}

\abstract{%
The International Lattice Datagrid (ILDG) is a federation of several
regional grids.  Since most of these grids have reached production
level, an increasing number of lattice scientists start to benefit from
this new research infrastructure.  The ILDG Middleware Working Group has
the task of specifying the ILDG middleware such that interoperability
among the different grids is achieved.
In this paper
we will present the architecture of the ILDG middleware and describe what
has actually been achieved in recent years.
Particular focus is given to interoperability and security issues.
We will conclude with a short overview on issues which we plan to address
in the near future.
}

\FullConference{The XXV International Symposium on Lattice Field Theory\\
		 July 30 - August 4 2007\\
		 Regensburg, Germany}

\begin{document}

\section{Introduction}

The goal of the International Lattice Datagrid (ILDG) is to establish
an international grid infrastructure which provides the means for longterm
storage and global sharing of data produced during compute-intensive
lattice Quantum Chromodynamics (QCD) simulations.

For data sharing and data procurement the availability of metadata is
a key requirement.  With common standards being developed, implemented
and used in practice it becomes possible to make metadata publicly
available such that all pieces of information have a well-defined
semantic meaning.
Within ILDG, the agreed standard for describing scientific metadata is
encapsulated in a family of XML schema, called QCDml.
For an overview and status report on the definition of QCDml
see~\cite{ildg-mdwg}.

The lattice QCD community, like many other research communities, is not in
the position to globally develop, deploy and operate a common middleware
stack.  Instead, application specific software has to be built on top
of existing middleware frameworks (e.g., gLite or the Globus Toolkit)
and grid infrastructures (like LCG/EGEE or the Open Science Grid). On the
global level this requires the realisation of a concept of grid-of-grids.
Making the various grid services interoperable continues to a major
challenge. Two strategies are employed to face the interoperability
challenge. Firstly, common grid standards are adopted, whenever possible.
An example is the Storage Resource Manager protocol (SRM) \cite{srm}, a
protocol evolving to an open standard for grid middleware to communicate
with site specific storage fabrics. Secondly, interface services are
defined and implemented. These services are implemented as web-services
for which a set of operations has been standardised by a
Webservice Description Language (WSDL) document,
a behavioural specification and the specification of test suites. For
instance, the ILDG file catalogue interface service allows to interface
to different file catalogues which are being used, e.g.~the
LCG File Catalogue (LFC) or the Globus Replica Location Service (RLS).

The following regional grids have been put into operation:
CSSM (Australia) \cite{cssm},
Latfor Datagrid (LDG) (continental Europe) \cite{ldg},
JLDG (Japan) \cite{jldg},
UKQCD grid (UK) \cite{digs}, and
USQCD (US) \cite{usqcd}.

%-------------------------------------------------------------------------------
\section{ILDG middleware functionality}

\begin{figure}[t]
\begin{center}
\includegraphics[scale=0.6]{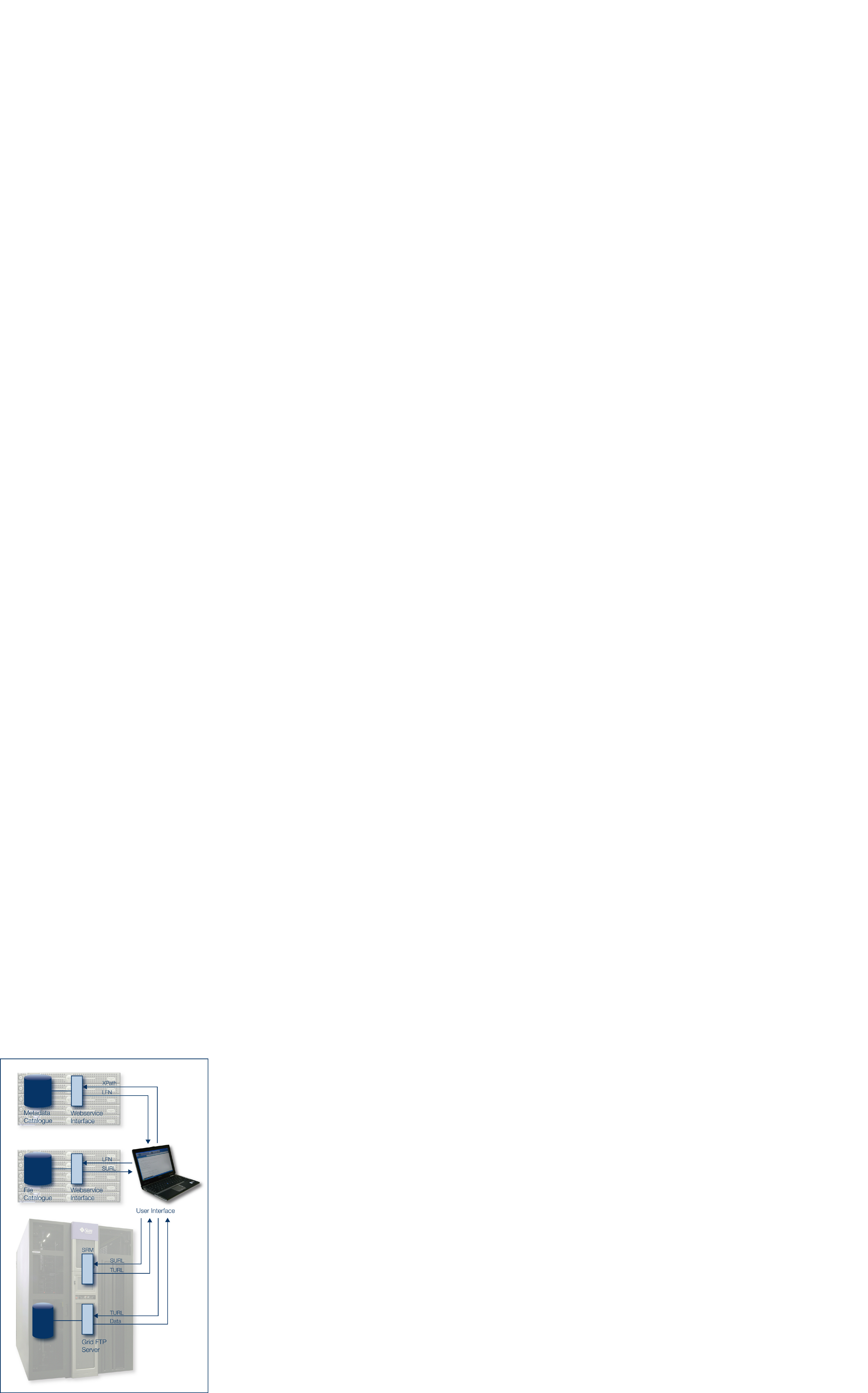}
\end{center}
\caption{\label{fig:arch}%
Simplified overview on the middleware architecture. On the left-hand
side the different services metadata catalogue (MDC), file catalogue
(FC) and Storage Element (SE) are shown (top to bottom)
which are deployed in all of the ILDG regional grids.
The arrows indicate the operations initiated by the client.}
\end{figure}

The currently deployed services allow to query for and to download
metadata as well as to download the data.\footnote{Note that within the
regional grids more operations, in particular data upload operations,
are available.} Typically, a user will start by performing
a query for available ensembles. The ensembles are described by
ensemble XML documents stored in a Metadata Catalogue (MDC). Each
ensemble can be addressed by a globally unique URI, the so-called
\texttt{markovChainURI}. The unique identifier for each configuration and
configuration metadata document is a logical filename, \texttt{dataLFN}.
(See~\cite{ildg-mdwg} for further details on the \texttt{markovChainURI}
and \texttt{dataLFN}.)
To actually download configurations that belong to a particular ensemble
the following operations have to be performed:
\begin{enumerate}\itemsep-1mm
\item Query the metadata catalogue (MDC) for all configurations which
      have a particular ensemble URI in common. The MDC will
      return a list of logical file names.
\item Query the file catalogue (FC) to identify all replicas of a file
      identified by its logical filename \texttt{dataLFN}. The FC will
      return a list of Site Uniform Resource Locators (SURL).%
      \footnote{An SURL is a kind of Uniform Resource Identifier (URI).}
\item If the scheme of the SURL is equal to 'srm' a Storage Resource
      Manager will
      have to be queried to obtain the actual transfer URL (TURL).
      Otherwise, SURL and TURL are identical.
\item Download the data from a Storage Element (SE) using the TURL.
\end{enumerate}
This sequence of operations and the involved services are schematically
shown in Fig.~\ref{fig:arch}. Note that a user does not have to care about
each of these steps as client tools will typically process a sequence of
these operations by default.

Operations needed to upload and modify data are currently not standardised
within ILDG and are hence regional grid specific.

%-------------------------------------------------------------------------------
\section{Middleware services}

The \emph{Metadata Catalogue} (MDC) is a database which stores the ensemble
and configuration XML documents. Each of the regional grids has deployed
such a service together with a webservice interface that provides at least the
following operations standardised by ILDG:
\begin{description}
\item[\texttt{doEnsembleURIQuery}]
Queries the MDC for ensemble metadata documents which match a given XPath
expression. For each matching document the corresponding
ensemble URI is returned.

\item[\texttt{doConfigurationLFNQuery}]
Similar to \texttt{doEnsembleURIQuery} but for configuration metadata
documents. A list of logical file names (LFN) is returned.

\item[\texttt{getEnsembleMetadata}]
Return ensemble metadata document for a given ensemble URI.

\item[\texttt{getConfigurationMetadata}]
Return configuration metadata document for a given LFN.
\end{description}
Several web-portals have been deployed by the regional grids which
allow to query all the MDCs. Furthermore, EPCC has developed the
ILDG Browser Client. This java-based browser allows to construct an
XPath query in an intuitive way and to submit this query to any of the
MDCs \cite{digs}.

The \emph{File Catalogue} (FC) stores for each logical file name
at least one Site Uniform Resource Locator (SURL). Such an SURL identifies a
copy of a data file (e.g.~gauge configuration) on any of the storage
elements. For each data file more than one replica may exist.
Different file catalogues are being used by the various
regional grids. Therefore, a webservice interface has been defined
and an interface service has been implemented as a joined effort by
DESY and EPCC. The interface service receives queries via the
standardised webservice interface, forwards them to the regional grid
specific FC and finally returns its response via the webservice
interface back to the client. So far only one interface operation is
available. The \texttt{getURL} operation takes
a list of LFNs on input and returns for each of them all registered SURLs.
All regional grids have deployed and tested this service.
A complication of the FC interface service results from authentication
being mandatory for some file catalogues. If this is the case then the
interface service has to able to act on behalf of the user. Or to put
it in different words: The user client has to delegate its credentials
to the interface service. (See below for more details on the credential
delegation service.)

The real data is stored on so-called Storage Elements (SE) where the
data can be found using its SURL. A large variety of SEs
configurations exist. A simple SE may consist of a single server with an
attached disk array. A large SE may comprise a whole set of file servers
which are used to stage the data stored in a tape back-end system (HSM).
Typically, the latter kind of SE will provide a Storage Resource
Manager (SRM) interface \cite{srm}.\footnote{SRM v1 is currently used,
but it is expected that the SEs will be upgraded to SRM
v2.2 in the near future.} The SRM protocol allows the client and server
to negotiate the transport protocol and to notify the client about the
server which will eventually provide the data.  This feature gives
the SE the freedom to dynamically assign a file server on which the
data is being staged depending, e.g., on its current load.  For final
data transport the ILDG middleware architecture allows for GridFTP
and http. This flexibility at the server side requires somewhat fatter
clients as ILDG download clients must be able to communicate with an SRM
interface and support all transport protocols.

We started to test interoperability of the services and verified that
from any of the regional grids download operations from SEs
operated by any of the regional grids is indeed possible. A prototype
client, called \texttt{ildg-get}, has been implemented. This allows to
download metadata from any of the metadata catalogues, to query the file
catalogues for replica of a particular file and, finally, to download the
file from an SE.

\begin{figure}[t]
\begin{center}
\includegraphics[scale=0.43]{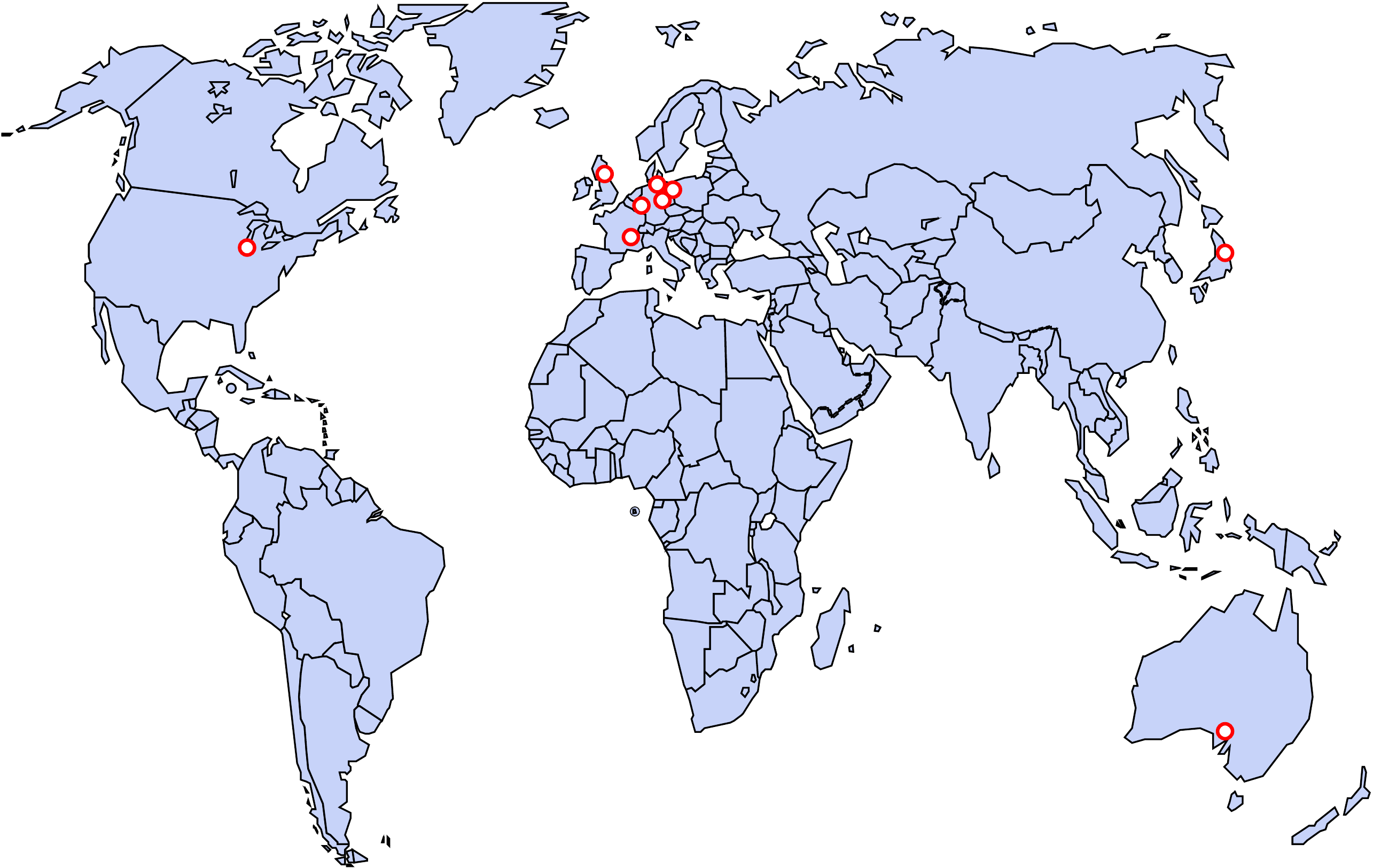}
\end{center}
\vspace*{-3mm}
\caption{ILDG storage elements currently in operation.}
\vspace*{-2mm}
\end{figure}

%-------------------------------------------------------------------------------
\section{Security and VO management}

The security infrastructure employed for ILDG is common to all major
grid middlewares. Before a secure operation is started user and service
perform a mutual \emph{authentication}. Both sides have to have a pair of
private key and public X.509 certificate.
The private key is confidential and
not transmitted via the network. The public certificate is signed by a
trusted Certificate Authority (CA). The server (client) can check whether
the private key/public certificate pair matches and whether
the user (host) certificate has been signed by a trusted CA. The later
is done by comparing the certificate with the root certificates published
by the CAs and which have to be installed both on the server and the client.

Trusting the CA means that we rely on the CAs to fulfil certain standards
when providing certificates to users, e.g.~they have to ensure that the
user did identify himself. To reduce the risk of user (host)
credentials being compromised the life-time of certificates are limited.
Furthermore, the CAs regularly publish lists of revoked certificates,
so-called Certificate Revocation Lists (CRL). Both, client and server
always check that the CRL is not out-of-date.
In ILDG it has been agreed to trust certificates
issued by any of the Certificate Authorities (CA) of the International
Grid Trust Federation (IGTF) \cite{igtf}.

To actually obtain access to any of the ILDG resources a user has
to join the \emph{Virtual Organisation} (VO) ILDG. Registration with
and management of the VO is done via the
\emph{Virtual Organization Membership Registration Service} (VOMRS)
\cite{vomrs}.
%A user which wants to join the VO he authenticates with the VOMRS
%service and perform the following operations:
%%
%\begin{itemize}
%\item Select a institution (i.e.~regional grid) to which he/she
%      belongs to.
%\item Select a representative who will decide on the applicant's
%      request to join the VO.
%\item Select a group to which it wants to belong to.
%\item Accept the VO Usage Rules as defined in the ILDG VO Policy
%      document \cite{vopolicy}.
%\end{itemize}
A user which wants to join to the VO has to perform the following
steps:
\vspace*{-1mm}
\begin{itemize}\itemsep-1mm
\item Load certificate into a web-browser and establish an authenticated
      connection to the VOMRS service.
\item In Phase I, the user fills in a registration form and nominates a
      regional grid (institution) plus representative, to approve their
      application.
\item After email address has been verified the candidate is invited to
      proceed to phase II of the registration procedure
      during which he/she can apply
      for joining a group and has to accept 
      the VO Usage Rules as defined in the ILDG VO Policy
      document \cite{vopolicy}.
\end{itemize}

There are 2 or 3 representatives for each of the regional
grids. They will decide only on those applications coming from the
corresponding regional grid. They are responsible for approving requests
according to the VO Policy document which, e.g., means that they have
to verify that the applicant is a scientist performing research in
lattice QCD.

The VOMRS service propagates VO membership information to a VOMS
service from where this information can be retrieved by each of the
regional grids. All regional grids automatically synchronise their
local user mappings with this central VO user management system.
Once a representative approved the application for VO membership
this new user will automatically become known to all services in
all the regional grids.

For each of the regional grids a corresponding group has been defined.
The information about group membership
can be used by the regional grid services for authorisation,
e.g.~to restrict access to data to members of that particular
regional grid. Only the owner and managers of a group are allowed to add
VO members to that particular group.

Particular care is needed when credentials are delegated by the client to
an interface service. For the FC interface service we use a credential
delegation service which has become part of the gLite middleware
\cite{delegation}. If a client requests a credential to be delegated
the server generates a private key and a certificate request. The latter
is sent to the client for being signed. After this public certificate
has been returned, the server can use it to authenticate on behalf of the
user with other servers. Note that the private key is never transmitted
via the (potentially insecure) network. For security reasons
the maximum time of validity of the delegated credentials is limited to
12 hours. Once the credential has expired the procedure of
credential delegation has to be repeated.

%-------------------------------------------------------------------------------
\section{Conclusions and outlook}

Significant progress has been made since ILDG was initially proposed
in 2001. Various regional grids have been deployed for production
use. This infrastructure is being used by lattice researchers all
around the world to make an increasing amount of data available.
An overview on the current usage has been given by C.~DeTar at
this conference \cite{carleton}.

Admittedly, a user who wants to gain access to the grid for the first time
has to make substantial efforts. The user has to
\vspace*{-1mm}
\begin{itemize}\itemsep-1mm
\item Install grid client software.
\item Identify a Certificate Authority (or one of its Registration
      Authorities) and request a certificate.
\item Join the Virtual Organisation.
\end{itemize}
We nevertheless observe that these efforts start to pay-off for an increasing
number of people being involved in lattice QCD research. In particular large
physics collaborations where tasks of generating gauge field configurations
and calculating observables on these configurations have been divided between
members distributed over different sites and countries start to benefit
from this new research infrastructure.

Still a lot of work needs to be done to improve the usability of this infrastructure.
Issues which the ILDG Middleware Working Group plans to address in the near
future include:
\vspace*{-1mm}
\begin{itemize}\itemsep-1mm
\item More efforts are needed to improve functionality and operation
      of all regional grids.
\item Reliable operation of all regional grid services should be improved by
      putting a monitoring system in place.
\item The available functionality of the services standardised by
      ILDG should be extended, in particular, to allow for replication of
      data beyond the regional grid boundaries. This would be relevant
      for physics collaborations which comprise of members in different
      regional grids. Furthermore, it would allow to optimise access time
      to data which would otherwise only be available from storage elements
      where the bandwidth of the network connecting client and server is
      poor.
\end{itemize}

%-------------------------------------------------------------------------------
\section*{Acknowledgements}

We are indebted to all institutions and people that helped to build-up the
regional grid infrastructures and provide their resources.
DP acknowledges the support from C.~Iezzi for preparing this presentation.

%-------------------------------------------------------------------------------


\begin{thebibliography}{99}
  \bibitem{ildg-mdwg}
    T.~Yoshi\'e et al. [ILDG Metadata Working Group],
    \emph{Marking up lattice QCD configurations and ensembles},
    \pos{PoS(LATTICE 2007)161}.

  \bibitem{srm}
    \emph{Storage Resource Manager protocol rev.~2.2}
    (\url{http://sdm.lbl.gov/srm-wg/doc/SRM.v2.2.pdf}).

  \bibitem{cssm}
    CSSM grid (\url{http://cssm.sasr.edu.au/ildg/}).

  \bibitem{ldg}
    LatFor Datagrid (\url{http://www-zeuthen.desy.de/latfor/ldg/}).

  \bibitem{jldg}
    Japan Lattice Datagrid (\url{http://www.jldg.org/lqa/index.html}).

  \bibitem{digs}
    QCDgrid/DiGS (\url{http://www.gridpp.ac.uk/qcdgrid/}).

  \bibitem{usqcd}
    USQCD grid (\url{http://usqcd.jlab.org/}).

  \bibitem{vomrs}
    Registration site for the ILDG virtual organisation:
    \url{https://grid-voms.desy.de:8443/vo/ildg/vomrs/}.

  \bibitem{igtf}
    International Grid Trust Federation (\url{http://www.gridpma.org}).

  \bibitem{vopolicy}
    ILDG Middleware Working Group, \emph{ILDG VO Policy} (Rev.~1.0).

  \bibitem{delegation}
    A.~McNab and S.~Kaushal, \emph{The GridSite Proxy Delegation Service},
    UK e-Science All Hands Conference, Nottingham, September 2006.

  \bibitem{carleton}
    C.~DeTar,
    \emph{Sharing Lattices Throughout the World: An ILDG Status Report},
    \pos{PoS(LATTICE 2007)009}.
\end{thebibliography}
\end{document}